\documentclass[10pt,conference]{IEEEtran}
\usepackage{graphicx}
\usepackage{xcolor,colortbl}
\definecolor{Gray}{gray}{0.85}
\usepackage{makecell}
\usepackage{xtab,booktabs}
\usepackage{tabularx}
\usepackage[skip=0.5\baselineskip]{caption}

\usepackage[colorinlistoftodos]{todonotes}

\usepackage{enumitem}
\usepackage{hyperref}
\usepackage{algorithm}
\usepackage{multirow}
\usepackage[noend]{algpseudocode}
\DeclareUnicodeCharacter{FB01}{fi}
\usepackage{authblk}

\usepackage{xspace}
\newcommand{\simname}{$\mathit{bin^{2}\!sim}$\xspace}

\newcommand{\approach}{\textsl{LibRARIAN}\xspace}
\newcommand{\oss}{\emph{OSSPolice}\xspace}
\newcommand{\giflib}{\emph{GIFLib}\xspace}

\newcommand{\gp}{{Google Play}\xspace}

\newcolumntype{C}[1]{>{\centering\arraybackslash}p{#1}}
\newcolumntype{L}[1]{>{\raggedright\arraybackslash}p{#1}}
\newcolumntype{R}[1]{>{\raggedleft\arraybackslash}p{#1}}
\newcolumntype{G}[1]{>{\columncolor{lgray}\centering\arraybackslash}p{#1}}

\usepackage{framed}
\newcounter{finding}[section]
\newenvironment{finding}{\refstepcounter{finding}
\vspace{-1mm}
\framed
\noindent \textbf{Finding~\thefinding:}
}
{
\endframed
\vspace{-1mm}
}

\newcommand{\vulapps}{{$53/200$}\xspace}
\newcommand{\vulappsPerc}{{$26.5\%$}\xspace}
\newcommand{\vulappsTillnow}{{$14$}\xspace}
\newcommand{\timetofixpatch}{{$528.71 \pm 40.20$}\xspace}
\newcommand{\timetoreleasepatch}{{$54.59 \pm 8.12$}\xspace}

\newcommand{\appvers}{{$7{,}678$}\xspace}
\newcommand{\totallibs}{{$66{,}684$}\xspace}

\newcommand{\gtlibs}{{$46$}\xspace}
\newcommand{\gtversions}{{$904$}\xspace}
\newcommand{\correctrate}{{$91.15\%$}\xspace}
\newcommand{\incorrectrate}{{$8.85\%$}\xspace}

\begin{document}
\title{Too Quiet in the Library: An Empirical Study of Security Updates in Android Apps' Native Code}

\author[$^\star$]{Sumaya Almanee}
\author[$^\star$]{Arda Ünal}
\author[$^\dagger$]{Mathias Payer}
\author[$^\star$]{Joshua Garcia}
\affil[$^\star$]{University of California Irvine,  \{salmanee, unala, joshug4\}@uci.edu}
\affil[$^\dagger$]{EPFL, mathias.payer@nebelwelt.net}

\maketitle
\begin{abstract}
Android apps include third-party native libraries to increase performance and to
reuse functionality.  Native code is directly executed from apps through the
Java Native Interface or the Android Native Development Kit. Android developers
add precompiled native libraries to their projects, enabling their use.
Unfortunately, developers often struggle or simply neglect to update these
libraries in a timely manner. This results in the continuous use of outdated
native libraries with unpatched security vulnerabilities years after patches
became available.

To further understand such phenomena, we study the security updates in native libraries in the most popular 200 free apps on \gp from Sept. 2013 to
May 2020. A core difficulty we face in this study is the identification of
libraries and their versions. Developers often rename or modify libraries,
making their identification challenging.
We create an approach called \approach (\textbf{LibRA}ry ve\textbf{R}sion
Identific\textbf{A}tio\textbf{N}) that accurately identifies native libraries
and their versions as found in Android apps based on our novel similarity metric
\simname. \approach leverages different features extracted from libraries based
on their metadata and identifying strings in read-only sections.

We discovered \vulapps popular apps (\vulappsPerc) with vulnerable versions with
known CVEs between Sept. 2013 and May 2020, with \vulappsTillnow of those apps remaining
vulnerable.  We find that app developers took, on average, \timetofixpatch
days to apply security patches, while library developers release a security
patch after \timetoreleasepatch days---a $10$ times slower rate of update.
\end{abstract}

\section{Introduction}

Third-party libraries are convenient, reusable, and form an integral part of
mobile apps. Developers can save time and effort by reusing already implemented
functionality.  Native third-party libraries are prevalent in Android
applications (``apps''), especially social networking and gaming apps. These two
app categories---ranked among the top categories on \gp---require special
functionality such as 3D rendering, or audio/video encoding/decoding
\cite{ffmpeg,libvpx,libvorbis,libopus,speex}. These tasks tend to be
resource-intensive and are, thus, often handled by native libraries to improve
runtime performance.

The ubiquity of third-party libraries in Android apps increases the attack
surface~\cite{seoFLEXDROIDEnforcingInApp2016,
sunNativeGuardProtectingAndroid2014} since host apps expose vulnerabilities
propagated from these libraries \cite{hpe_report_2016, sonatype_report}. Another
series of previous work has studied the outdatedness and updateability of
third-party \emph{Java libraries} in Android apps \cite{derrKeepMeUpdated2017,
backesReliableThirdPartyLibrary2016}, with a focus on managed code of such apps
(e.g., Java or Dalvik code). However, these previous studies do not consider
\emph{native libraries} used by Android apps.

We argue that security implications in native libraries are even more critical
for three main reasons. First, app developers add native libraries but neglect
to update them. The reasons for this may include concerns over regressions
arising from such updates, prioritizing new functionality over security,
deadline pressures, or lack of tracking library dependencies and their security
patches. This negligence results in outdated or vulnerable native libraries
remaining in new versions of apps.  
Second, native libraries are susceptible to memory vulnerabilities (e.g., buffer
overflow attacks) that are straight-forward to exploit.
Third, and contrary to studies from almost 10 years
ago~\cite{enckStudyAndroidApplication2011, zhouHeyYouGet2012}, native libraries
are now used pervasively in mobile apps.
To illustrate this point, we analyzed the top 200 apps from \gp between Sept. 2013 and May 2020. We obtained the version
histories of these apps from \textit{AndroZoo}~\cite{allixAndroZooCollectingMillions2016} totaling
\appvers versions of those 200 top free apps. From these
apps, we identified \totallibs native libraries in total with an
average of $11$ libraries per app and a maximum of $141$ for one version of \textit{Instagram}.

To better understand the usage of third-party native libraries in Android apps
and its security implications, we conduct a longitudinal study to identify
vulnerabilities in third-party native libraries and assess the extent to which
developers update such libraries of their apps. In order to achieve this, we
make the following research contributions:

\begin{itemize}[leftmargin=*,nosep]
\item We construct a novel approach, called \approach (\textbf{LibRA}ry
ve\textbf{R}sion Identific\textbf{A}tio\textbf{N}) that, given an unknown
binary, identifies (i) the library it implements and (ii) its version.
Furthermore, we introduce a new similarity-scoring mechanism for comparing
native binaries called \simname, which utilizes $6$ features that enable
\approach to distinguish between different libraries and their versions. The
features cover both metadata and data extracted from the libraries. These
features represent elements of a library that are likely to change between major, minor, and
patch versions of a native library.

\item We conduct a large-scale, longitudinal study that tracks security
vulnerabilities in native libraries used in apps over $7$ years. We build a
repository of Android apps and their native libraries with the $200$ most
popular free apps from \gp totaling \appvers versions gathered between the dates
of Sept. 2013 and May 2020. This repository further contains \totallibs native
libraries used by these \appvers versions.

\end{itemize}

Prior work \cite{mingBinSimTracebasedSemantic,
liaoMobileFindrFunctionSimilarity2018, huBinMatchSemanticsbasedHybrid2018,
ALAM2017230} has measured the similarity between binaries. However, these
approaches identify semantic similarities/differences between binaries at the
\emph{function-level}, with the goal of identifying malware. \approach,
orthogonally, is a syntactic-based tool which computes similarity between two
benign binaries (at the file-level) with the goal of identifying library
versions with high scalability.

We utilize \approach and our repository to study (1) \approach's accuracy and effectiveness,
(2) the prevalence of vulnerabilities in native libraries in the top 200 apps, and 
(3) the rate at which app developers apply patches to address vulnerabilities in
native binaries.
The major findings of our study are as follows:

\begin{itemize}[leftmargin=*,nosep]
\item For our ground truth dataset which contains \gtlibs known
  libraries with \gtversions versions, \approach correctly identifies
  \correctrate of those library versions, thus achieving a high
  identification accuracy.

\item To study the prevalence of vulnerabilities in the top 200 apps in \gp, we use \approach to examine $53$ apps with vulnerable versions and
known CVEs between Sept. 2013 and May 2020. \vulappsTillnow of these apps remain
vulnerable and contain a wide-range of vulnerability types---including denial of
service, memory leaks, null pointer dereferences, or divide-by-zero errors. We
further find that libraries in these apps, on average, have been outdated for
$859.17 \pm 137.55$ days. The combination of high severity and long exposure of
these vulnerabilities results in ample opportunity for attackers to target these
highly popular apps.

\item To determine developer response rate of applying security fixes,
  we utilize \approach to analyze $40$ apps, focusing on popular
  third-party libraries (those found in more apps) with known CVEs
  such as \emph{FFmpeg}, \giflib, \emph{OpenSSL}, \emph{WebP},
  \emph{SQLite3}, \emph{OpenCV}, \emph{Jpeg-turbo}, \emph{Libpng}, and
  \emph{XML2}, between Sept.  2013 and May 2020.

   We find that app developers took, on average,
\timetofixpatch days to apply security patches, while library developers release
a security patch after \timetoreleasepatch days---a $10$ times slower rate of
update. These libraries that tend to go for long periods without being patched
affect highly popular apps with billions of downloads and installs.

\item We make our dataset, analysis platform, and results available
  online to enable reusability, reproducibility, and others to build
  upon our work \cite{librarian}.

\end{itemize}

\section{\approach}\label{sec:approach}

\begin{figure}[h]
\centering
\includegraphics[width=0.5\textwidth]{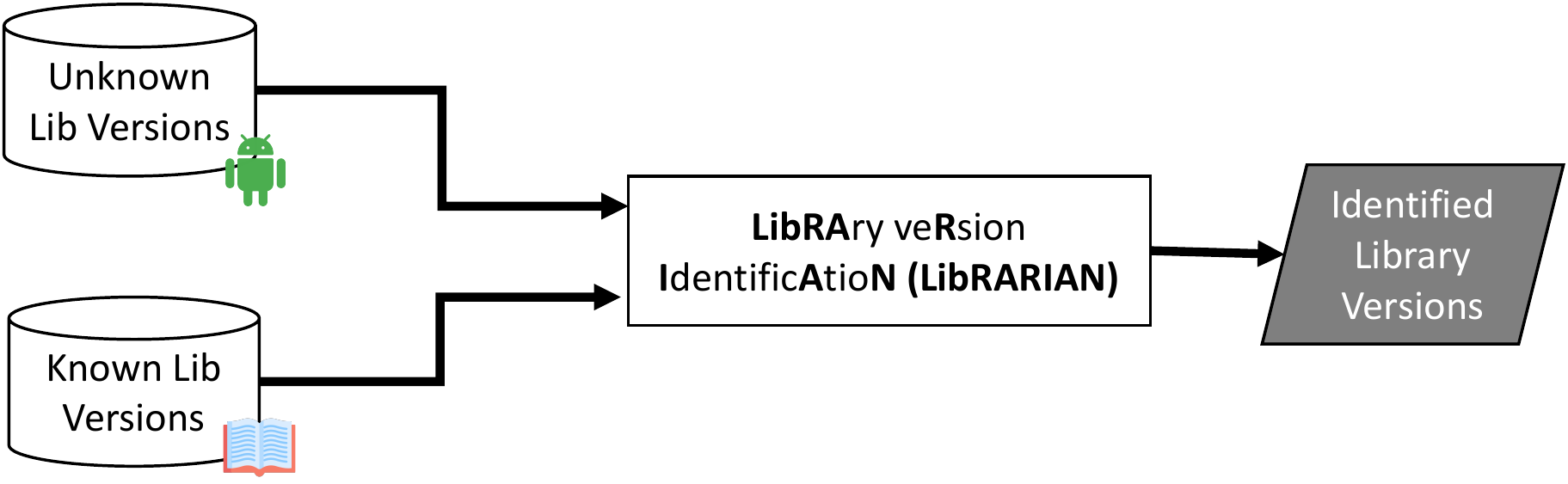}
\caption{\approach identifies versions of native binaries from Android
  apps by using our \simname similarity-scoring technique to compare
  known (ground-truth dataset) and unknown versions of native binaries.
} 
\label{fig:approach}
\end{figure}

\autoref{fig:approach} shows the overall workflow of
\approach. \approach identifies unknown third-party native libraries
and their versions (\emph{Unknown Lib Versions}) by (1) extracting
features that distinguish major, minor, and patch versions of
libraries that are stable across platforms regardless of underlying
architecture or compilation environments; (2) comparing those features
against features from a ground-truth dataset (\emph{Known Lib
  Versions}) using a novel similarity metric, \simname; and (3)
matching against strings that identify version information of
libraries extracted from \emph{Known Lib Versions}, which we refer to
as \textit{Version Identification Strings}. In the remainder of this
section, we describe each of these three major steps of \approach.

\subsection{Feature Vector Extraction}\label{sec:fv}

Our binary similarity detection is based on the extraction of features
from binaries combining both metadata found in Executable and Linkable
Format (ELF) files as well as identifying features in different binary
sections of the library.  All shared libraries included in Android
apps are compiled into ELF binaries. Like other object files, ELF
binaries contain a symbol table with externally visible identifiers
such as function names, global symbols, local symbols, and imported
symbols. This symbol table is used (1) during loading and linking and
(2) by binary analysis tools~\cite{GnuOrg} (e.g., \texttt{objdump},
\texttt{readelf}, \texttt{nm}, \texttt{pwntools}, or
\texttt{angr}~\cite{shoshitaishviliSOKStateArt2016}) to infer
information about the binary.

\begin{table*}[h!]
\centering
\begin{tabular}{ |C{1cm}|p{3.5cm}|p{10cm}| } 
\hline
 \rowcolor{Gray}
  \textbf{Feature Type} & \textbf{Name} & \textbf{Definition} \\ [0.5ex] 
  \hline 
\multirow{5}[0]{*}{\rotatebox[origin=c]{90}{\textbf{Metadata}}} &
  Exported Globals  &  Externally visible variables, i.e., they can
be accessed externally.  \\ & 
  Imported Globals   & Variables from other libraries that are used in this
library.   \\ &
  Exported Functions  & Externally visible functions, i.e., functions 
that can be called from outside the library. \\ &
  Imported Functions & Functions from other libraries that are used in this
library. \\ &
  Dependencies  & The library dependencies that are automatically
                  loaded by the ELF object    \vspace{5pt} \\  \hline
  \multirow{1}[0]{*}{\rotatebox[origin=c]{90}{\textbf{Data}}} &           
  Version Identification Strings  & Flexible per-library version strings
(e.g., ``libFoo-1.0.2'' matched to strings in the \textit{.rodata} section of an ELF object)   \\
\hline
\end{tabular}
\caption{List of features \approach extracts from native binaries of Android apps along with their type and definition.}

\label{tab:fv}
\end{table*}  

To distinguish between different libraries and their versions, we need
to identify \emph{differencing features}. To that end, we define a set
of six features inherent to versions and libraries.  Five features
represent ELF metadata, these features are used to compute the
similarity score between two binaries as described in \autoref{sec:sc},
hence, we refer to these features as \textit{Metadata
  Features}. Orthogonally, we leverage strings extracted from the
\textit{.rodata} section of an ELF object, which we refer to as
\textit{Version Identification Strings}. This feature complements the
similarity score from the first set of features. We either use it to
verify the correctness of the version or as a fallback if the
similarity to existing binaries in our ground-truth dataset is low
(see \autoref{sec:versionid}).

\autoref{tab:fv} shows the list of all \approach features. The
features include: (i) five \textit{Metadata Features} based on
exported and imported functions, exported and imported globals, and
library dependencies; and (ii) one \textit{Data Feature} which is
applied as a second factor to either substitute the \textit{Metadata
  Features}, in case the reported similarity score is low, or to
confirm the reported score. These 6 features represent the code
elements of a library that would be expected to change based on a
versioning scheme that distinguishes major, minor, and patch versions
of a library. Furthermore, these features are stable across platforms
regardless of the underlying architecture or compilation
environments. We did not include code features (e.g., control-flow and
data-flow features) as they are extremely volatile and change between
compilations and across architectures. Binary similarity matching is a
hard open problem: While recent work has made progress regarding
accuracy~\cite{eschweilerDiscovREEfficientCrossArchitecture2016,
  xuPatchBasedVulnerability2020,mingBinSimTracebasedSemantic,liaoMobileFindrFunctionSimilarity2018,huBinMatchSemanticsbasedHybrid2018,ALAM2017230,hemelFindingSoftwareLicense2011,duanIdentifyingOpenSourceLicense2017},
the majority of algorithms have exponential computation cost relative
to the code size and are infeasible for large-scale studies.

We built a dataset of heuristics by inspecting the binaries in our
ground-truth dataset. We developed scripts to process the data in the
\textit{.rodata} sections extracted during feature processing and
search for unique per-library strings that contain version
information. For example, \textit{FFmpeg} version info is found when
applying the regex \verb/ffmpeg-([0-9]\.)*[0-9]/ or
\verb/FFmpeg version([0-9]\.)*[0-9]/. \autoref{tab:versionfeatures}
shows our list of extracted version heuristics.
Each version heuristic can be produced automatically by
constructing regular expressions from strings in \textit{.rodata}
sections of binaries in our ground-truth dataset. For example, if the
string ``\textit{libFoo-1.0.2}'' is found in version 1.0.2 of \textit{libFoo}, \approach
uses a regular expression replacing the numeric suffix of the string
with an appropriate pattern (e.g., \verb/libFoo-[0-9]+(\.[0-9])*/).

We deliberately exclude any metadata or identifying strings for symbols that are
volatile across architectures or build environments like compiler
version, relocation information (and types), or debug symbols. \approach's
accuracy results in \autoref{sec:eval:rq1:ia} demonstrate that our selected set
of features suffice to distinguish between different versions of libraries. 

\begin{table*}[h!]
\centering
\begin{tabular}{|l|l|}
  \rowcolor{Gray}\hline
  \textbf{Library Name} & \textbf{Extracted Heuristics} \\\hline
  	Jpeg-turbo&\verb/Jpeg-turbo version 1(\.[0-9]{1,})*/\\
  	FFmpeg&\verb/ffmpeg-([0-9]\.)*[0-9]|FFmpeg version ([0-9]\.)*[0-9]/\\
  	Firebase&\verb/Firebase C++ [0-9]+(\.[0-9])*/\\
  	Libavcodec&\verb/Lavc5[0-9](\.[0-9]{1,})/\\
 	Libavfilter&\verb/Lavf5[0-9](\.[0-9]{1,})/\\
 	Libpng&\verb/Libpng version 1(\.[0-9]{1,})*/\\
	Libglog&\verb/glog-[0-9]+(\.[0-9])*/\\
	Libvpx&\verb/WebM Project VP(.*)/\\
	OpenCV&\verb/General configuration for OpenCV [0-9]+(\.[0-9])*|opencv-[0-9]+(\.[0-9])*/\\
	OpenSSL&\verb/openssl-1(\.[0-9])*[a-z]|^OpenSSL 1(\.[0-9])*[a-z]/\\
	Speex&\verb/speex-(.*)/\\
  	SQLite3&\verb/^3\.([0-9]{1,}\.)+[0-9]/\\
	Unity3D&\verb/([0-9]+\.)+([0-9]+)[a-z][0-9]|Expected version:(.*)/\\
	Vorbis&\verb/Xiph.Org Vorbis 1.(.*)/\\
	XML2&\verb/GITv2.[0-9]+(\.[0-9])/\\
  \hline
  \end{tabular}
  \caption{Heuristics used to search for unique per-library strings that contain version information}
  \label{tab:versionfeatures}
\end{table*}

The implementation leverages
angr's~\cite{shoshitaishviliSOKStateArt2016} ELF parser which already
is platform independent. Our extraction platform recovers all metadata
from the ELF symbol tables and, if available, searches for string
patterns in the comment and read-only sections. Our filters remove
platform specific information and calls to standard libraries (e.g.,
C++ ABI calls, vectors, or other data structures). The current
implementation covers x86-64, x86, ARM, and ARM64 binaries---which are
all platforms we observed in our evaluation.
We accommodate for architecture differences in two ways:
First, we remove architecture noise in feature vectors (e.g., symbols that are
only used in one architecture); and second, we collect, if available, binaries
for the different architectures.

The feature extraction compiles all recovered information as a dictionary into a
JSON file. The dictionary contains arrays of strings for each of the features
mentioned above plus additional metadata to identify the library and
architecture.

\subsection{Similarity Computation}\label{sec:sc}

\approach's similarity computation, which we refer to as \simname,
leverages the five \textit{Metadata Features} when computing the similarity
scores between an app binary and our ground-truth dataset. \simname is
based on the \textit{Jaccard coefficient},
and is used to determine the similarity between feature vectors.
\simname allows \approach to account for addition or removal of
features between different libraries and versions.  Given two binaries
$b_1$ and $b_2$ with respective feature vectors $FV_{1}$ and $FV_{2}$,
\simname computes the size of the intersection of $FV_{1}$ and
$FV_{2}$ (i.e., the number of common features) over the size of the
union of $FV_{1}$ and $FV_{2}$ (i.e., the number of unique features):

\vspace{-3ex}
\begin{equation}\label{eq:1}
bin^2sim(FV_{1}, FV_{2}) = \frac{|\;FV_{1}\;\cap\;FV_{2}\;|}{|\;FV_{1}\;\cup\;FV_{2}\;|} \quad\in\;[0,1]
\quad
\end{equation}
\vspace{-3ex}

The similarity score is a real number between 0 and 1, with a score of
1 indicating identical features, a score of 0 indicating no shared
features between the two libraries, and a fractional value indicating a
partial match.
Due to the volatility of the similarity score, filtering noise such as
platform-specific details as mentioned in the previous section is essential for
the accuracy of our approach.


\approach counts an unknown library instance from \emph{Unknown Lib
  Versions} as matching a known library version if its \simname
is above 0.85. This threshold was determined experimentally and works
effectively as our evaluation will demonstrate (see
\autoref{sec:eval}). If \simname results in the same value above
the threshold for multiple known binaries, \approach tries obtaining
an exact match between one of the known binaries and the unknown
binary by using their hash codes to determine the unknown binary's
version.



    



A low similarity score might result from modifications made by app developers to
the original third-party library which results in the removal or addition of
specific features. From our experience, removal of features from the original
library is common among mobile developers and is likely driven by the need to
reduce the size of the library and the app as much as possible. For example, we
observed that the WebP video codec library is often deployed without encoding
functionalities to reduce binary size.  Some size optimization techniques
require choosing needed modules from a library and leaving the rest, stripping
the resulting binary, and modifying build flags. Another factor that reduces
similarity as measured by the Jaccard coefficient is that certain architectures
tend to export more features as compared to others. For instance, 32-bit
architectures such as armeabi-v7a and x86 export more features compared to
arm64-v8a and x86\_64.

\subsection{Version Identification Strings}\label{sec:versionid}

For libraries where \approach reports low similarity scores (e.g.,
some libraries like \textit{RenderScript} or \textit{Unity} only
export a single function \footnote{These libraries are ``stripped''
  and hide all functionality internally. The single exported function
  takes a string as parameter which corresponds to the target function
  and they dispatch to internal functionality based on this string.}),
these five features fail to provide sufficient information about the
underlying components in a library. If libraries only export one or a
few functions, the similarity metrics have a hard time distinguishing
between different libraries. We therefore extend the features with
strings that uniquely identify the library. Such strings are often
version strings. Based on extracted flexible per-library heuristics
from our ground-truth dataset (see \autoref{tab:versionfeatures}), we
heuristically identify exact library versions and increase overall
accuracy. For libraries with high similarity scores, we use these
library heuristics to confirm the correct version.


To identify binaries with low similarity scores, we leverage 
\textit{Version Identification Strings}, which is the set of extracted
per-library version strings. For example, say a library version
$lv$ extracted from app $a$ had a similarity score of 0.3 when
compared with \textit{OpenCV-2.4.11} using \textit{Metadata
  Features}. Given the low score, we search the \textit{Version
  Identification Strings} feature for specific keywords such as
\texttt{General configuration for OpenCV *.*.*} or
\texttt{opencv-*.*.*}. Where the asterisk represents the versioning
scheme of \textit{OpenCV} library.

Our feature extraction process logs all strings (arrays of more than 3
ASCII printable characters ending with a 0 byte) from the
\textit{.rodata} section alongside the other features. As
libraries commonly have large amounts of read-only string data that
frequently changes, we cannot use this data directly as a feature (due
to the low overlap resulting in low similarity).  By processing the
\textit{.rodata} from our ground-truth dataset and clustering the data,
we extract common version identifiers and version strings. We then
translate them into regular expressions that allow us to match
versions for different libraries.



\section{Evaluation}\label{sec:eval}

To assess the prevalence of vulnerable native libraries for Android,
we answer the following three research questions:
\begin{enumerate}[leftmargin=*,label=\bfseries
  RQ\arabic*:,nosep,wide=0pt]
\item \emph{Accuracy and effectiveness of \approach.} Can \approach
  accurately and effectively identify versions of native libraries?
  How does \approach compare against state-of-the-art native-library
  version identification? How effective are \approach's feature types
  at identifying versions of native libraries?
\item \emph{Prevalence of outdated libraries.} How prevalent are vulnerabilities in native
  libraries of Android apps?
\item \emph{Patch response time.} After a vulnerability is reported for a third-party library,
  how quickly do developers apply patches?
  \end{enumerate}

To supplement the aforementioned RQs, we conducted a detailed case study on a
vulnerable app (\autoref{sec:eval:exploit}), providing practical insight into
vulnerabilities in third-party libraries and possible exploits.

  To answer these research questions, we analyze the top 200 apps in
  \gp over several years.  We track the version history of these apps
  from AndroZoo \cite{allixAndroZooCollectingMillions2016}, a large
  repository of over 11 million Android apps. Our repository contains
  app metadata including the app name, release dates, and native
  binaries.
Note that \gp unfortunately restricts lists to 200 apps.
Overall, we collected \appvers instances, where each instance is a version of
the $200$ top apps from \gp.

We determined that $145$ out of $200\;(72.50\%)$ of the distinct apps
in our repository contain at least one native library, i.e., $5{,}852$
out of $7{,}678\;(76.21\%)$ of the total apps in our database. There
are a total of \totallibs libraries in the form of \textit{.so} files,
i.e., shared library files, in our repository with an average of $11$
libraries per app and a maximum of $141$ for one version of
\textit{Instagram}. In fact, \textit{Instagram}---for which we
collected $184$ versions since Dec. 2013---contains a total of
$6{,}677$ \textit{.so} files.

We run \approach on a machine with 2 AMD EPYC 7551 32-Core CPUs and 512GB of RAM
running Ubuntu 18.04. The average number of features in the extracted feature
vectors is $2{,}116.86$ features. Some outliers such as \textit{libWaze} and
\textit{libTensorflow} reach up to $79{,}581$ features. This shows that the set
of third-party native libraries in our repository is diverse, some of them are
very complex and offer a large number of functionalities. 
Generating feature vectors is quick and generally takes a few seconds per
library. The most complex library, \textit{libTensorflow} takes $4$ min and $38$
sec to analyze. We found that, out of 7,253 binaries for which \approach
inferred their versions, the average runtime for library version detection is
118.19 seconds---with a minimum of 97 seconds and a maximum of 224 seconds.

\subsection{RQ1: Accuracy and Effectiveness}\label{sec:eval:rq1}

To determine if \approach accurately and effectively identifies native
library versions from Android apps, we assess \approach in three
scenarios. For the first scenario, we compare its accuracy with \oss,
the state-of-the-art technique for identifying versions of native
binaries for Android apps. For the second scenario, we assess
\approach on a larger dataset for which \oss could not be applied and,
thus, evaluate \approach's accuracy independently of other tools.  In
the third scenario, we assess the effectiveness of \approach's feature
types at identifying versions of native libraries.

\subsubsection{Comparative Analysis}\label{sec:eval:rq1:comp}

\oss uses source code to build an index that allows it to identify versions of
binaries.
\oss measures the similarity between strings extracted from binaries
and features found directly in source repositories. Unlike \approach,
\oss relies on comparing binaries with source code, resulting in an
overly large feature space which, in turn, makes \oss susceptible to
falsely identifying any binary containing a library as exactly
matching that library.  For example, \oss falsely identifies
\textit{MuPDF} and \textit{OpenCV} as matching \textit{Libpng} because
those two libraries include \textit{Libpng} in their source code
\cite{duanIdentifyingOpenSourceLicense2017}.

We repeatedly contacted the \oss authors to obtain a fully-working
version of their tool, but unfortunately they did not provide us their
non-public data index or sufficient information to reproduce their
setup.
As a result, we performed a comparative analysis between \approach and
\oss based on the published \oss
numbers~\cite{duanIdentifyingOpenSourceLicense2017}.

The ground-truth dataset in the \oss evaluation contains a total of
$475$ binaries (out of which $67$ are unique) extracted from $104$
applications collected by
F-Droid~\cite{fdroid}. 
%
%
\approach correctly identified $63/67$ $(94\%)$ unique binaries in the
\oss dataset, improving accuracy by 12\% compared to the accuracy
reported by \oss ($82\%$) which correctly identified $55/67$
libraries. \oss has lower accuracy because it misidentifies
reused libraries (as described above) and it relies
on simple syntactic features (e.g., string literals and
exported functions) while our feature vectors extract additional
features---such as imported functions, exported and imported global
variables, and dependencies that uniquely identify different versions
of binaries. These additional features were a major factor in the
superior accuracy of \approach compared to \oss.

\approach did not identify $4$ binaries because the library functions are
dispatched from a single function and do not contain identifying version
information that was readily available. Hence, our extracted features fail to
provide sufficient information about the underlying components in the
library. Nevertheless, \approach significantly reduces the number of
binaries that need to be manually inspected.

Lastly, it is important to reiterate that these results are only
compared against the dataset used in the \oss paper but without us
being able to replicate or reuse \oss, due to key elements of the tool
being unavailable.



\begin{finding}
  \label{find:rq1_oss}
  \approach achieves a 12\% improvement in its accuracy compared to
  \oss on the 67 unique binaries in \oss's dataset. Unlike \oss,
  \approach obtains this improvement without relying on source code,
  which may not be available for all libraries and results in an
  unnecessarily larger feature space.
\end{finding}

\subsubsection{Independent Accuracy}\label{sec:eval:rq1:ia}

We further assess \approach's accuracy on a larger and more recent set
of library versions than those found in \oss's dataset. To that end,
we manually collect a set of binaries with known libraries and
versions (\emph{Known Lib Versions} in \autoref{fig:approach}) and
compare the inferred libraries and versions to the known ones to
determine \approach's accuracy. We build our dataset based on
libraries used in common Android apps.

\textbf{Experiment Setup}. We first manually locate the pre-built
binaries of libraries to serve as ground truth.  To that end, we use
readily available auxiliary data such as keywords found in feature
vectors, binary filenames, and dependencies.  Once we identify
potential targets, we retrieve the pre-built binaries of all versions
and architectures, if possible.

There are a variety of distribution channels where app developers can obtain
third-party binaries. We obtained such binaries from official websites, GitHub,
and Debian repositories. The binaries with known libraries and versions contain
\gtlibs distinct libraries with a total of \gtversions versions and an average of $19$
versions per library.

\textbf{Results}.
\approach correctly identified the versions of $824/904$ (\correctrate)
libraries in our ground truth: $553/904$ ($61.17\%$) of these library
versions have unique feature vectors; $15.16\%$ of the these libraries
contain the exact version number in the strings literals; and the
remaining $14.82\%$ of library versions are distinguished using hash
codes to break ties between \simname values of binaries. 

Misidentification occurs in \incorrectrate of library versions, where
the largest equivalence class contains 4 library versions. This
usually occurs for consecutive versions---minor or micro revisions
(e.g., 3.1.0 and 3.1.1). These minor or micro revisions generally fix
small bugs and do not change, add, or remove exported
symbols. Although \approach cannot pinpoint the exact library version
in this case, \approach significantly reduces the search space for
post analysis to a few candidate versions.

\begin{finding}
  \label{find:rq1_indie}
  \approach correctly identifies 824 of 904 (91.15\%) library versions
  from 46 distinct libraries, making it highly accurate for
  identifying the native libraries and versions. For misidentified
  library versions, \approach reports a slightly different version.

\end{finding}

\subsubsection{Feature Effectiveness}\label{sec:eval:rq1:ef}

To assess the effectiveness of \textit{Metadata Features},
\textit{Version Identification Strings}, and their combination at
inferring binaries, we computed the extent to which each feature is
capable of inferring binaries in our repository. To that end, any
binary whose library and version can be inferred with a \simname above
0.85 as described in \autoref{sec:approach} counts as an inferred
binary. We found that $37.42\%$ of binaries in our repository are
inferable by \textit{Version Identification Strings} only. $45.29\%$
of the remaining binaries are inferable using only the five
\textit{Metadata Features} mentioned in \autoref{sec:fv}, while the
remaining $17.29\%$ are inferred using both \textit{Metadata Features}
and \textit{Version Identification Strings}. This indicates that not
all libraries have the version information encoded directly in the
strings. Having a combination of both \textit{Metadata Features} and
\textit{Version Identification Strings} is crucial to increase the
number of inferred binaries.

We further aimed to assess the extent to which each of the five
\textit{Metadata Features} contribute to computing \simname in order
to assess each of their individual effectiveness. Recall from
\autoref{sec:sc} that our matching algorithm leverages five features
when computing the similarity scores between an app binary and our
ground-truth dataset. \autoref{tab:fv-con} lists these feature along
with their contribution factor, i.e., the average percentage each one
of these features contribute to the total similarity score. To calculate the
contribution factor ($contrib_{f}$) of a feature $f$, we first calculate the
similarity score taking all five features into account ($score_{all}$). We then
calculate the similarity score of each feature separately ($score_{f}$). For
each feature, we find $contrib_{f}=score_{f}/score_{all}$, which is the
percentage each $f$ contributes to the total similarity score. As shown in
\autoref{tab:fv-con}, \textit{Exported Functions} contributes the most when
computing \simname (\autoref{eq:1}), i.e., 58.25\% of the matching features are
\textit{Exported Functions}, followed by \textit{Imported Functions}
contributing 32.98\%, \textit{Dependencies}, \textit{Exported Globals}, and
finally \textit{Imported Globals} contributing less overall. Still, these five
features sometimes manage to uniquely identify a library and are therefore
included as they, overall, improve the similarity score.
Recall that \textit{Version Identification Strings} is not taken into account
when computing the similarity score between binaries. 



\begin{finding}
  \label{find:rq1_features}
  $37.42\%$ of binaries are inferable using \textit{Version
  Identification Strings}, $45.29\%$ are inferable using
\textit{Metadata Features}, and \textit{17.29\%} are inferable using
both feature types. \textit{Exported Functions} and \textit{Imported
  Functions} account for the overwhelming majority of effectiveness of
\textit{Metadata Features}, contributing $58.25\%$ and $32.98\%$,
respectively.

\end{finding}

\begin{table}[h!]
\centering
\begin{tabular}{ |C{1cm}|p{3.5cm}|R{1.5cm}| } 
\hline
 \rowcolor{Gray}
  \textbf{Feature Type} & \textbf{Name} &\textbf{Contribution Factor}  \\ [0.5ex] 
  \hline 
\multirow{5}[0]{*}{\rotatebox[origin=c]{90}{\textbf{Metadata}}} &
Exported Globals  &  3.32\%  \\ & 
Imported Globals  & 1.06\%  \\ &
Exported Functions &58.25\% \\ &
Imported Functions & 32.98\% \\ &
Dependencies  & 4.39\%  \\
\hline
\end{tabular}
\caption{List of features \simname extracted from native binaries of
  Android apps along with their type and overall contribution factor,
  which measures the average percentage each feature contributes to
  the total similarity score}
\label{tab:fv-con}
\end{table}  

\subsection{RQ2: Prevalence of Vulnerable Libraries}\label{sec:eval:rq2}

To study the prevalence of vulnerabilities in native libraries, we
need to identify their exact versions. To that end, we leverage
\approach to identify potential library versions from our
repository. Once the versions are identified, we investigate the
extent to which native libraries of Android apps are vulnerable and
remain vulnerable.

\textbf{Experiment Setup}. We infer the correct version of $7{,}253$ binaries ($10.87\%$ of the total binaries in our Android
repository) using \approach. Due to the highly
time-consuming nature of the manual collection of ground-truth binaries, we limit
ourselves to libraries that (i) are found in a greater number of apps
(more than $10$ apps) and (ii) have known \emph{CVEs}. As a result, an
overwhelming majority of the remaining binaries in our dataset have
either no known CVEs or affect very few apps, making them an
unsuitable choice for applying an expensive manual analysis for
studying this research question.

\begin{table*}[h]
  \centering
  \footnotesize
\begin{tabular}{|p{1.5cm}|R{1.2cm}|p{10cm}|R{1.2cm}|R{1.2cm}|} 
\rowcolor{Gray}
\hline
\textbf{Lib Name} & \textbf{No. Vul Lib Vers} & \textbf{Vul Lib Vers} & \textbf{No. Apps} & \textbf{No. Apps Still Vul} \\ 
\hline
OpenCV&5&2.4.1, 2.4.11, 2.4.13, 3.1.0, 3.4.1&21&7\\
WebP&3&0.3.1, 0.4.2, 0.4.3&11&1\\
GIFLib&2&5.1.1, 5.1.4&15&1\\
FFmpeg&9&2.8, 2.8.7, 3.0.1, 3.0.3, 3.2, 3.3.2, 3.3.4, 3.4, 4.0.2&8&1\\
Libavcodec&9&55.39.101, 55.52.102, 56.1.100, 56.60.100, 57.107.100, 57.17.100, 57.24.102, 57.64.100, 57.89.100&10&0\\
Libavformat&3&55.19.104, 56.40.101, 57.71.100&3&0\\
Libavfilter&3&3.90.100, 4.2.100, 5.1.100&1&0\\
Libavutil&3&52.48.101, 52.66.100, 54.20.100&2&0\\
Libswscale&3&2.5.101, 3.0.100, 4.0.100&5&1\\
Libswresample&2&0.17.104, 1.1.100&1&0\\
SQlite3&7&3.11.0, 3.15.2, 3.20.1, 3.26.0, 3.27.2, 3.28.0, 3.8.10.2&7&2\\
XML2&1&2.7.7&3&1\\
OpenSSL&22&1.0.0a, 1.0.1c, 1.0.1e, 1.0.1h, 1.0.1i, 1.0.1p, 1.0.1s, 1.0.2a, 1.0.2f, 1.0.2g, 1.0.2h, 1.0.2j, 1.0.2k, 1.0.2m, 1.0.2o, 1.0.2p, 1.0.2r, 1.1.0, 1.1.0g, 1.1.0h, 1.1.0i, 1.1.1b&13&3\\
Jpeg-turbo&2&1.5.1, 1.5.2&3&0\\
Libpng&7&1.6.10, 1.6.17, 1.6.24, 1.6.34, 1.6.37, 1.6.7, 1.6.8&5&1\\
\hline
\end{tabular}
\caption{A list of libraries with reported \emph{CVEs} found in our
  repository along with the number of distinct apps that were affected by a
  vulnerable library and the number of distinct apps containing a vulnerable
  version till now.}
\label{tab:vulnerablelibs}
\end{table*}

\textbf{Results}. We found that, out of $7{,}253$ binaries for which
we inferred their versions, $3{,}674$ were vulnerable libraries
($50.65\%$) affecting \vulapps distinct apps. \vulappsTillnow new releases of
these distinct apps remain vulnerable at the time of submission. The
complete list of libraries with reported \emph{CVEs} between
Sept. 2013 and the writing of this paper can be found in
\autoref{tab:vulnerablelibs}. As for the number of apps affected by
vulnerable libraries, our results show that $53$ distinct apps have
been affected by a minimum of $1$ vulnerable library and a maximum of
$16$ vulnerable libraries covering dates between Sept 2013 and May
2020.

\begin{finding}
\label{rq3_prevalence} 
53 of the 200 top apps on Google Play (26.5\%) were plagued by a
vulnerable library over approximately six years and 8 months (i.e.,
between Sept. 2013 and May 2020).  \vulappsTillnow of those apps still include a
vulnerable binary, i.e., 7\% of the top 200 apps on Google Play, even
at the time at which we collected apps for this study and are, on average,
outdated by $859.17 \pm 137.55$ days. As a result,
vulnerable native libraries play a substantial role in exposing
popular Android apps to known vulnerabilities.
\end{finding}

We emailed app developers since February 2020 to inform them that their apps
continue to use a vulnerable library. We urged them to take an action (i.e.,
remove or replace such libraries) or at least provide some justification as to
why such libraries are not updated.  Our investigation is ongoing. While several
app developers already updated their apps to remove the vulnerable library,
many updates are still outstanding. Some of the replies we received simply
blame other library developers. For example, we heard back from
Discord that the vulnerable lib is a dependency of another third-party library
used in Discord (Fresco): ``Until Fresco fixes this, however, we are not able to
address this in our app''.

Four libraries were particularly prevalent in terms of the number of
vulnerable versions they contain (i.e., \textit{OpenSSL}), the number
of apps they affect (i.e., \textit{OpenCV} and \giflib), or the length
of time during which the library remained vulnerable (i.e.,
\textit{XML2} in \textit{Microsoft XBox SmartGlass}). \emph{OpenSSL}
has the largest number of vulnerable versions ($22$ in total) included
in $13$ distinct apps. $3$ apps: Amazon Alexa, Facebook Messenger and Norton Secure VPN
still include vulnerable versions of \emph{OpenSSL}.\

\emph{OpenCV} and \giflib affect the most apps. \emph{OpenCV} has the
largest number of affected apps with a total of $21$ apps where $7$
recent apps still have a vulnerable instance of \emph{OpenCV}. 
Most applications do not include OpenCV directly but indirectly
through the dependencies of \texttt{card.io} which enables card
payment processing but comes with the two outdated versions (2.4.11
and 2.4.13) of both \texttt{opencv\_core} and \texttt{opencv\_imgproc}.  Following
\emph{OpenCV} in the number of affected apps is \giflib, which has two
vulnerable versions found in a total of $15$ distinct apps, $1$ app is still
affected.

One vulnerable version of \emph{XML2} (2.7.7) was found in $35$
versions of \emph{Microsoft XBox SmartGlass} and the library was not
updated for 6 years---still remaining vulnerable up to the writing of
this paper. This particular case is notable due to the extremely long
amount of time the library had been vulnerable and remained
vulnerable.

To examine the affects of vulnerable libraries on apps further, we
list popular apps and the reported CVEs they expose their users
to. \autoref{tab:10vulapps} shows 10 out of \vulappsTillnow popular
apps that are using at least one library with a reported CVE at the
time of our app collection. We discuss four of these apps in more
detail in the remainder of this section.

\emph{Facebook Messenger}, which has a download base of over $500$M (the largest
in this list), contains \emph{OpenSSL-1.1.0}, which is vulnerable since Sept.
2016. This vulnerable library contains multiple memory leaks which allows an
attacker to cause a denial of service (memory consumption) by sending large OCSP
(Online Certificate Status Protocol) request extensions. 

\emph{Amazon Kindle}, an app that provides access to an electronic
library of books---with a total of more than $100$M installs, uses two
vulnerable libraries: \emph{XML2-2.7.7} and
\emph{Libpng-1.6.7}. \emph{XML2-2.7.7} contains a variant of the
``billion laughs'' vulnerability which allows attackers to craft an XML
document with a large number of nested entries that results in a
denial of service attack. \emph{XML2-2.7.7} is vulnerable
since Nov. 2014 and continues to be used in recent versions of the
app. \emph{Libpng-1.6.7} has a NULL pointer dereference
vulnerability. This vulnerability was published 6 years ago under
\emph{CVE-2013-6954} and it remains unchanged in recent releases of
\emph{Amazon Kindle}.

\emph{DoorDash}, a food delivery app with more than $10$M installs
includes \giflib-5.1.4 which was reported vulnerable over 8 months
ago. A malformed GIF file triggers a division-by-zero exception in the
\emph{DGifSlurp} function in \giflib versions prior to 5.1.6. This
vulnerable library remains unchanged up to now.

\emph{Target} uses \emph{OpenCV-2.4.11} as a dependency of
\texttt{card.io} which enables card payment processing. This version
of \emph{OpenCV} was announced vulnerable in Aug. 2017 yet remains
unchanged in these apps. 

\renewcommand{\arraystretch}{1.25}
\begin{table}[h!]
\centering
\begin{tabular}{|p{2.5cm}p{4cm}R{1cm}|}
\rowcolor{Gray}\hline
\textbf{App Name} & \textbf{Vulnerable Libs} &\textbf{No. Installs} \\\hline
Amazon Alexa&OpenSSL-1.0.2p, SQlite3-3.27.2&10M+\\
Amazon Kindle&Libpng-1.6.7, XML2-2.7.7&100M+\\
Amazon Music &FFmpeg-4.0.2&100M+\\
DoorDash &GIFLib-5.1.4&10M+\\
Facebook Messenger &OpenSSL-1.1.0&500M+\\
Grubhub &OpenCV-2.4.1&10M+\\
Sam's Club&OpenCV-2.4.1&10M+\\
SUBWAY&OpenCV-2.4.1&5M+\\
Norton Secure VPN &OpenSSL-1.1.1b&10M+\\
Target&OpenCV-2.4.11&10M+\\\hline
\end{tabular}
\caption{10 out of \vulappsTillnow popular apps from \gp which include a vulnerable library
that remained unchanged.}
\label{tab:10vulapps}
\end{table}  


\begin{finding}
\label{find:rq2_vultypes}
These four apps showcase that these vulnerabilities are wide-ranging
involving denial of service, memory leaks, or null pointer
dereferences. The high severity and long exposure time of
these vulnerabilities results in ample opportunity for attackers to
target these highly popular apps.
\end{finding}

\subsection{RQ3: Rate of Vulnerable Library Fixing}\label{sec:eval:rq3}

To determine the vulnerability response rate, we identify the duration
between (1) the release time of a security update and (2) the time at
which app developers applied a fix either by (i) updating to a new
library version or (ii) completely removing a vulnerable library.
Recall that we collected the previous versions of the top 200 apps
from \gp.  Moreover, we inferred the library versions from $7{,}253$
libraries using \approach.  Given the histories of apps and inferred
library versions we can track the library \emph{life span} per
app---i.e., the time at which a library is added to an app and when it
is either removed or updated to a new version in the app.


To this end, we analyzed $40$ popular apps with known vulnerable
versions of \emph{FFmpeg}, \giflib, \emph{OpenSSL}, \emph{WebP},
\emph{SQLite3}, \emph{OpenCV}, \emph{Jpeg-turbo}, \emph{Libpng}, and \emph{XML2},
between Sept. 2013 and May 2020. We exclude apps that removed a library before a \emph{CVE} was associated with it and apps containing
libraries that are vulnerable up to the writing of this paper. We
obtained the date at which a library vulnerability was found; when a
security patch was made available for the library; and the time at
which either the library was updated to a new version or
removed. \autoref{tab:timetofixAppLib} shows all the combinations of
apps and vulnerable libraries.

\begin{finding}
  \label{find:rq3_slow_patching}
On average, library developers release a security patch
after \timetoreleasepatch days from a reported \emph{CVE}. App developers
apply these patches, on average, after \timetofixpatch days from
the date an update was made available---which is about $10$ times
slower than the rate at which library developers release security
patches. 
\end{finding}

\autoref{find:rq3_slow_patching} reveals that many popular Android
apps expose end-users to long vulnerability periods, especially
considering that library developers released fixed versions much
sooner. This extreme lag between release of a security patch for a
library and the time at which an app developer updates to the patched
libraries, or just eliminates the library, indicates that, at best, it
is (1) highly challenging for developers to update these kinds of
libraries or, less charitably, (2) app developers are highly negligent
of such libraries.

\begin{table}[h]
  \centering
  \footnotesize
\begin{tabular}{|p{2cm}p{1.9cm}p{1.4cm}R{0.6cm}R{0.8cm}|}
\rowcolor{Gray}\hline
\textbf{App Name}&\textbf{Vul Lib Version}& \textbf{Vul Announced} & \textbf{TTRP (Days)}&\textbf{TTAF (Days)}\\\hline
Xbox&XML2-2.7.7&2014-11-04&12&1956\\
Apple Music&XML2-2.7.7&2014-11-04&12&1704\\
TikTok&GIFLib-5.1.1&2015-12-21&87&1429\\
Zoom Meetings&OpenSSL-1.0.0a&2010-08-17&91&1323\\
Amazon Alexa&OpenSSL-1.0.1s&2016-05-04&12&1086\\
Amazon Kindle&Libpng-1.6.34&2017-01-30&330&1019\\
StarMaker&FFmpeg-3.2&2016-12-23&4&1001\\
eBay&OpenCV-2.4.13&2017-08-06&41&905\\
Fitbit&SQlite3-3.20.1&2017-10-12&12&902\\
Uber &OpenCV-2.4.13&2017-08-06&41&830\\
Snapchat&SQlite3-3.20.1&2017-10-12&12&670\\
Discord &GIFLib-5.1.1&2015-12-21&87&665\\
Lyft &OpenCV-2.4.11&2017-08-06&41&662\\
Twitter&GIFLib-5.1.1&2015-12-21&87&457\\
Instagram&FFmpeg-2.8.0&2017-01-23&2&267\\
\hline
\end{tabular}
\caption{Combinations of 15 apps and particular vulnerable library
  versions they have contained, the date the vulnerability was
  publicly disclosed (\emph{Vul announced}), the period between vulnerability
  disclosure and patch availability in days (i.e. Time-to-Release-Patch (\emph{TTRP})), and the total number of days elapsed before a fix was made (i.e. Time-to-Apply-Fix (\emph{TTAF}))}
\label{tab:timetofixAppLib}
\end{table}
\vspace{1ex}

Developers applied security patches for vulnerable libraries at a rate
as slow as 5.4 years, in the case of \emph{Xbox}, and as fast as 267
days for \emph{Instagram}, where a vulnerable version of \emph{FFmpeg} was
removed in that amount of time. In order to determine what type of fix
was applied by a developer, we checked the next app version where a
vulnerable library was last seen. We found that developers either kept
the library but updated to a new version, removed a vulnerable
version, or removed all native libraries in an app. In the next
paragraphs, we discuss five popular native libraries used in Android
apps that exhibit particularly slow fix rates: \emph{FFmpeg},
\emph{OpenSSL}, \giflib, \emph{OpenCV}, and \emph{SQLite3}.

Multiple vulnerabilities were found in versions 2.8 and 3.2 of
\emph{FFmpeg} in Dec. 2016 and Jan. 2017, respectively. The number of
days a security patch was released for these vulnerable library
versions is $4$ and $2$ days, respectively.  However, developers took
$267$ days to address vulnerabilities in \emph{Instagram}, and nearly
$3$ years to apply a fix in \emph{Starmaker}.

\emph{OpenSSL-1.0.0a} and \emph{OpenSSL-1.0.1s} were associated with \emph{CVE-2010-2939} and \emph{CVE-2016-2105} in Aug. 2010, and May 2016 of which \emph{OpenSSL} developers provided a security patch $91$ and $12$ days after. However, developers of \emph{Zoom} took $1{,}323$ days to apply a fix, while developers of \emph{Amazon Alexa} took $1{,}086$
days.

A heap-based buffer overflow was reported in \giflib-5.1.1 at the end of
2015. The results show that $3$ apps using this vulnerable version of \giflib have
an average time-to-fix, i.e., total number of days elapsed before a fix was
applied, of $850.33$ days ($2.3$ years), which is $10$ times slower.  This lag
time is particularly concerning since \giflib released a fix 87 days after the
vulnerable version. 

A fix to an out-of-bounds read error that was affecting \emph{OpenCV} through
version 3.3 was released $41$ days after the CVE was published. The vulnerable
versions of this library affects $3$ apps in total: \emph{Uber}, \emph{Lyft}, and
\emph{eBay}. \emph{OpenCV} has an average time-to-fix of $799$ days (i.e., $2$
years), which is $19$ times slower than the rate at which library developers
of \emph{OpenCV} release security patches.

\emph{SQLite3} released version 3.26.0, which fixes an integer
overflow found in all versions prior to 3.25.3. \emph{Snapchat} and
\emph{Fitbit} removed a vulnerable version of \emph{SQLite-3.20.1}
library $786$ days later.


\begin{finding}
  \label{find:rq3_slow_but_severe}
  The results for these five popular native libraries in Android apps
  show that it often takes years for app developers to update to new
  library versions---even if the existing version contains severe
  security or privacy vulnerabilities---placing millions of users at
  major risk.
\end{finding}

\begin{table}[h!]
\centering
\begin{tabular}{|p{2.4cm}|R{2.5cm}|R{1.5cm}|} 
\rowcolor{Gray}\hline
\textbf{App Name}&\textbf{Time-to-Apply-Fix (Days)}&\textbf{No. Installs}\\\hline
Apple Music&1704.00&50M+\\
Amazon Kindle&1019.00&100M+\\
eBay&905.00&100M+\\
Fitbit&902.00&10M+\\
Snapchat&844.00&1,000M+\\
Xbox&763.67&50M+\\
ZOOM Meetings&668.33&100M+\\
Lyft &662.00&10M+\\
Amazon Alexa&605.50&10M+\\
Uber&588.50&500M+\\\hline
\end{tabular}
\caption{Top 10 most negligent apps in terms of the average time to fix a vulnerable library}
\label{tab:timetofixApp}
\end{table}  

To further understand the consequences of outdated vulnerable
libraries, we calculated the average time-to-fix across all vulnerable
libraries per app. \autoref{tab:timetofixApp} lists the top 10 apps
with the most number of days a vulnerable library remained in an app
until a fix for the vulnerability was applied. \emph{Apple Music} had
the longest lag between the vulnerable library being introduced and
fixed, i.e., $4.66$ years. \emph{Uber} was the fastest at almost $589$
days. Individual apps had as few as over $10$ million installs and as
many as over a billion installs. Among the social-media apps,
\emph{Snapchat}, which has over 1 billion downloads and the largest
number of installs among the top 10 apps in
\autoref{tab:timetofixApp}, fixed its vulnerable libraries after $844$
days. These very long times to fix vulnerable libraries in highly
popular social-media apps places billions of users at high security
risk.

\begin{finding}
  \label{find:rq3_most_neglected_apps}
  The most neglected apps in terms of time to fix vulnerable native
  libraries range from 588.50 days to nearly five years, affecting
  billions of users and leaving them at substantial risk of having
  those libraries exploited. This finding emphasizes the need for
  future research to provide developers with mechanisms for speeding
  up this very slow fix rate.
\end{finding}


\autoref{tab:timetofixLib} lists the top 10 most neglected vulnerable libraries across all apps. \emph{XML2} is the most neglected library with an average time-to-fix of $5$ years; \emph{WebP} is the least neglected library with an average time-to-fix of $213.40$ days. Among these 10 libraries, the fact that it takes app developers $431.81$ days, on average, to update vulnerable
versions of \textit{OpenSSL} is particularly concerning due to its
security-critical nature.

\begin{table}[h!]
\centering
\begin{tabular}{|p{1.9cm}|R{2.4cm}|p{2.7cm}|} 
\rowcolor{Gray}\hline
\textbf{Lib Name}&\textbf{Time-to-Apply-Fix (Days)}&\textbf{Genre}\\\hline
XML2&1830.00&XML parser\\
Libpng&923.20&Codec\\
Jpeg-turbo&841.67&Codec\\
FFmpeg&720.90&Multimedia framework\\
OpenCV&635.27&Computer Vision\\
OpenSSL&431.81&Network\\
GIFLib&421.06&Graphis\\
SQlite3&369.29&RDBMS\\
WebP&213.40&Codec\\\hline
\end{tabular}
\caption{Top 10 most neglected vulnerable libraries in terms of the
  average time-to-fix}
\label{tab:timetofixLib}
\end{table}  

\begin{finding}
  \label{find:rq3_most_negelected_libs}
  Future research should focus on these highly neglected libraries as experimental
  subjects for determining methods to ease the burden of updating
  these libraries; running regression tests to ensure these updates do
  not introduce new errors; and repairing those errors, possibly
  automatically, when they do arise.
\end{finding}

\subsection{Exploitability Case Study}\label{sec:eval:exploit}

To demonstrate the exploitability of unpatched vulnerabilities in
third party apps, we carry out a targeted case study where we analyze
individual applications and create a proof-of-concept (PoC)
exploit. Our PoC highlights how these unpatched vulnerabilities can be
exploited by third parties when interacting with the apps.

\emph{XRecoder} allows users to capture screen videos, screen shots, and record
video calls. Furthermore, XRecoder provides video editing functionalities,
enabling users to trim videos and change their speed. This application uses
FFmpeg, an open-source video encoding framework that provides video and audio
editing, format transcoding, video scaling and post-production effects.

XRecoder embeds the FFmpeg library version 3.1.11, which is vulnerable to
\emph{CVE-2018-14394} (reported in July 2018).  FFmpeg-3.1.11 contains a
vulnerable function (\emph{ff\_mov\_write\_packet}) that may result in a
division-by-zero error if provided with an empty input packet. Hence, an
attacker can craft a WaveForm audio to cause denial of service. 

To assess whether this vulnerable function is reachable in XRecoder, we used
\emph{Radare2} \cite{r2} to replace the first instruction in the vulnerable
function with an interrupt instruction. We run the application after the latter
modification which consequently resulted in an app crash, i.e., allowing us to
trigger the vulnerability consistently.

\texttt{ff\_mov\_write\_packet} is called by multiple functions across two different
binaries (FFmpeg-3.1.11.so and the app-specific libisvideo.so) and two
different platforms (Dalvik and Native).  \texttt{av\_buffersink\_get\_frame},
one of the ancestors of \texttt{ff\_mov\_write\_packet}, is called by
\texttt{nativeGenerateWaveFormData} from the Dalvik-side. 

\section{Discussion:}\label{sec:discussion}
 
Findings in \emph{RQ2} (\autoref{sec:eval:rq2}) demonstrate that out of
$7{,}253$ binaries for which we inferred their versions, $3{,}674$ were
vulnerable libraries ($50.65\%$) affecting $53$ distinct apps between Sept. 2013
and May 2020.  This constitutes about \vulappsPerc of the top 200 apps on \gp. More
alarmingly, new releases of \vulappsTillnow distinct apps remain vulnerable even at the
time at which we collected apps for this study with an average outdatedness of
$859.17 \pm 137.55$ days. While we have informed app developers about the
outdated libraries in their apps, one interesting piece of follow-up work based
on this result is surveying Android app developers to determine the reason for
this extremely slow rate of fixing vulnerable native libraries in their apps.
Such a study can further assess what forms of support app developers would need
to truly reduce this slow rate of updating vulnerable library versions to ones
with security patches.

For RQ3 (\autoref{sec:eval:rq3}), we analyzed the speed at which developers
updated their apps to patched libraries and found that, on average, library
developers release a security patch after \timetoreleasepatch days from a reported
\emph{CVE}. While app developers apply these patches on average after $528.71
\pm 41.20$ days from the date an update was made available ($10$ times slower).
Recall that we only consider apps in these cases that actually ended up fixing
vulnerable native libraries. The results for RQ2 and RQ3 corroborate the need to
make app developers aware of the severe risks they are exposing their users to
by utilizing vulnerable native libraries. 


Overall, our results demonstrate the degree to which native libraries are
neglected in terms of leaving them vulnerable.
Unfortunately, our findings indicate that the degree of negligence of native
libraries is severe, while popular apps on \gp use native libraries
extensively with 145 out of 200 top free apps ($72.50\%$). Interesting future work
for our study includes uncovering the root causes of such negligence and means
of aiding developers to quickly update their native libraries. For example,
platform providers (e.g., Google) could provide mechanisms to automatically
update native libraries while also testing for regressions and possibly
automatically repairing them. Such an idea is similar to how Debian's
repositories centrally manage libraries and dependencies between applications
and libraries. Whenever a library is updated, only the patched library is
updated, the applications remain the same. The Android system would highly
profit from a similar approach of central dependency and vulnerability
management.

\section{Threats to Validity}\label{sec:threats}
\textbf{External validity}.  The primary external threat to validity
involves the generalizability of the data set collection and the
selection methodology. Recent changes in \gp limited the length of the
“top-apps” list to 200 items. Despite the restrictions imposed by \gp
(limiting our analysis to the top-200 apps), these apps (1) account
for the bulk of downloads and the largest user base on \gp and (2) are
generalizable to popular apps, thus having the largest impact.


The results from \emph{RQ1} show that \approach detects versions of
native libraries with high accuracy (\correctrate). The need to
compare against binaries with a known number of versions and libraries
(i.e., \textit{Known Lib Versions} in \autoref{fig:approach}) limits
\approach.  Specifically, misidentification of a library or its
version might occur when an unknown binary for which we are trying to
identify a library and version does not exist in \textit{Known Lib
  Versions}. In these cases, \approach identifies the unknown binary
as being the library and version closest to it according to \simname
that exists in \textit{Known Lib Versions}. One possible way of
enhancing \approach in such cases is to leverage supervised machine
learning, which may, at least, be able to identify if the library is
most likely an unknown major, minor, or patch version of a known
library.

\textbf{Internal validity}.  One internal threat is the accuracy of
timestamps in AndroZoo and its effect on the reported patch life cycle
findings. To mitigate this threat, we collected AndroZoo timestamps
over three months and correlated updates with \gp. We verified that
AndroZoo has a maximum lag of 9 days. This short delay is much smaller
than the update frequency of vulnerable apps. Furthermore, we verified
that using dates added to AndroZoo and version codes give us reliable
timestamps for earlier time periods.



\textbf{Construct validity}. One threat to construct validity is the
labeling of the libraries in our repository as vulnerable or not. To
mitigate this threat, we relied on the vulnerabilities reported by the
Common Vulnerabilities and Exposures database \cite{cve} which
contains a list of publicly known security vulnerabilities along with
a description of each vulnerability.

We conducted an exploitability case study of one vulnerable library in an app
\autoref{sec:eval:exploit}. For the remaining set of discovered vulnerable
libraries/apps, we verified that vulnerable native functions are exported and
that the library is loaded from the app/Dalvik-side. Performing a complete
analysis of exploitable/reachable native functions in Android is an interesting
but orthogonal research problem. Building a cross-language
control-flow/data-flow analysis to assess reachability of vulnerable native code
from the Dalvik code of an Android app is an open research problem, worthy of a
separate research paper: (1) recovering a binary CFG/DFG is currently unsound,
based on heuristics, and runs into state explosion and (2) conducting an
exploitability study of all vulnerable libraries/apps across our entire dataset
is infeasible due to the large amount of apps/libraries.

Another threat to validity is the possibility of developers manually patching
security vulnerabilities. To mitigate this threat to validity, we checked the
versions identified by \approach and found that \approach correctly identifies
an overwhelming majority of patch-level versions (61.21\%). For the patch-level
versions that \approach cannot distinguish as effectively, \approach makes
manual identification much easier, by significantly reducing the search space
for post analysis to only 3-4 candidate versions. Furthermore, based on the
results of our dataset, we believe that app developers are unlikely to manually
patch a library they do not maintain given that it already takes years for these
developers to simply update a library version.



\section{Related Work}\label{sec:related-work}
A series of work has demonstrated the importance of third-party
libraries for managed code of Android apps (i.e., Dalvik code) and
their security effects and implications \cite{derrKeepMeUpdated2017,
  backesReliableThirdPartyLibrary2016}. Derr et
al. \cite{derrKeepMeUpdated2017} investigated the outdatedness of
libraries in Android apps by conducting a survey with more than 200
app developers. They reported that a substantial number of apps use
outdated libraries and that almost $98\%$ of $17K$ actively used
library versions have known security vulnerabilities. Backes et
al. \cite{backesReliableThirdPartyLibrary2016} report, for managed
code-level libraries, that app developers are slow to update to new
library versions---discovering that two long-known security
vulnerabilities remained present in top apps during the time of their
study. None of these studies examined native third-party libraries in
Android apps nor did they look at the security  impact of vulnerable libraries
or whether these vulnerabilities are on the attack surface. 
\approach now explores the attack surface of native libraries, closing this
important gap and calling platform providers to action.

A wide variety of approaches have emerged that identify third-party
libraries with a focus on managed code.  These approaches employ
different mechanisms to detect third-party libraries within code
including white-listing package names
\cite{graceUnsafeExposureAnalysis2012,bookLongitudinalAnalysisAndroid2013a};
supervised machine learning
\cite{narayananAdDetectAutomatedDetection2014,liuEfficientPrivilegeDeEscalation2015};
and code clustering \cite{wangWuKongScalableAccurate2015,
  maLibRadarFastAccurate2016, liLibDScalablePrecise2017}. LibScout
\cite{backesReliableThirdPartyLibrary2016} proposed a different
technique to detect libraries using normalized classes as a feature
that provides obfuscation resiliency.

Some techniques identify vulnerabilities in native libraries by
computing a similarity score between binaries with known
vulnerabilities and target binaries of interest
\cite{gaoVulSeekerSemanticLearning2018}\cite{eschweilerDiscovREEfficientCrossArchitecture2016}.
VulSeeker \cite{gaoVulSeekerSemanticLearning2018} matches binaries
with known vulnerabilities using control-flow graphs and machine
learning.  Similarly, discovRE
\cite{eschweilerDiscovREEfficientCrossArchitecture2016} and BinXray \cite{xuPatchBasedVulnerability2020} matches binaries at the function
level.
Other techniques employ a hybrid technique such as
\emph{BinSim}\cite{mingBinSimTracebasedSemantic},
\emph{Mobilefinder}\cite{liaoMobileFindrFunctionSimilarity2018},
\emph{BinMatch}\cite{huBinMatchSemanticsbasedHybrid2018}, and \emph{DroidNative}
\cite{ALAM2017230}. These approaches identify semantic similarities/differences
between functions in binaries based on execution traces for the purpose of
analyzing/identifying malware. Unlike these tools, \approach focuses on benign
libraries with the goal of identifying their versions with high scalability.

Binary Analysis Tool (BAT) \cite{hemelFindingSoftwareLicense2011} and
\oss \cite{duanIdentifyingOpenSourceLicense2017} measure similarity
between strings extracted from binaries and features found directly in
source repositories. Unlike \approach, these approaches compare source
code with binaries, which introduces the issue of internal clones
(i.e., third-party library source code that is reused in the source
code of another library).  BAT and \oss rely on simple syntactic
features (e.g., string literals and exported functions).  \oss cannot
detect internal code clones, while \approach can, giving it superior
ability to identify versions of native libraries.  Furthermore, BAT
does not detect versions of binaries and was shown to have inferior
accuracy for computing binary similarity compared to \oss. Unlike
these tools, \approach extracts additional features---such as imported
functions, exported and imported global variables, and dependencies
that uniquely identify different versions of binaries. As shown in
\autoref{sec:eval:rq1:comp}, these additional features were a major
factor in the superior accuracy of \approach compared to \oss.

Other related empirical research studies the prevalence of vulnerable
dependencies in open source projects
\cite{carlsonOpenSourceVulnerability2019}, vulnerabilities in
WebAssembly binaries \cite{lehmannEverythingOldNew}, or investigates
the updatability of ad libraries in Android Apps
\cite{mojicaruizAnalyzingAdLibrary2016}. Other work such as
\cite{heDiversifiedThirdpartyLibrary2020,
  yuCombiningCollaborativeFiltering2017} present third party library recommendation tools for mobile apps.

Despite the existence of much previous work on survivability of vulnerabilities
in Android apps/libraries, such work has not conducted a large-scale
longitudinal study of native third-party libraries as we did in this paper.
Moreover, the survivability of vulnerabilities in non-native libraries are
significantly shorter compared to those reported in our results. While
survivability of vulnerabilities in native Android apps took, on average,
\timetofixpatch days in our study, prior work \cite{jsandpython,andoridOS} shows
that survival times of vulnerabilities in Python and Javascript are 100 days and
365 days, respectively. 50\% of vulnerabilities in npm packages were fixed
within a month, 75\% were fixed within 6 months only \cite{npm}.

None of this aforementioned related work has examined the prevalence of security
vulnerabilities in Android's native libraries or the time-to-fix for vulnerable
versions of such libraries. As a result, our work covers a critical attack
vector that has been ignored in existing research.

\section{Conclusion}\label{sec:conclusion}

Third-party libraries have become ubiquitous among popular apps in the official
Android market, \gp, with $145$ out of the $200$ top free apps on \gp
($72.50\%$) containing native libraries. These libraries are particularly
beneficial for handling CPU-intensive tasks and for reusing existing code in
general. Unfortunately, the pervasiveness of native third-party libraries in
Android apps expose end users to a large set of unpatched security
vulnerabilities.

To determine the extent to which these native libraries remain
vulnerable in Android apps, we study the prevalence of
native libraries in the top $200$ apps on \gp across $7{,}253$
versions of those apps. From these versions, we extracted \totallibs
native libraries. To identify versions of libraries, we constructed an
approach called \approach that leverages a novel similarity metric,
\simname, that is capable of identifying versions of native libraries
with a high accuracy---a \correctrate correct identification rate.

For vulnerabilities, we found $53$ apps with vulnerable versions with
known CVEs between Sept. 2013 and May 2020, with \vulappsTillnow of
those apps still remaining vulnerable until the end point of our
study. We find that app developers took, on average, \timetofixpatch
days to apply security patches, while library developers release a
security patch after \timetoreleasepatch days---a $10$ times slower
rate of update. 

\section{Data Availability}\label{sec:da}
Our dataset, analysis platform, and results are available online
\cite{librarian} for reusability and reproducibility purposes.

\section{Acknowledgement}\label{sec:ack}
This work was supported in part by award CNS-1823262 and CNS-1801601 from the
National Science Foundation and grant number 850868 from the ERC Horizon 2020 program.

\clearpage
\bibliographystyle{plain}
\bibliography{AndroLib}

%
%
%

\end{document}